\documentclass[12pt,a4paper]{article}
\usepackage[utf8]{inputenc}
\usepackage{units}
\usepackage{textcomp}
\usepackage{amsmath}
\usepackage{graphicx}
\usepackage{setspace}
\makeatletter

\pdfpageheight\paperheight
\pdfpagewidth\paperwidth

\@ifundefined{date}{}{\date{}}
\makeatother

\begin{document}

\title{Stirring, mixing, growing: microscale processes change larger scale
phytoplankton dynamics.}

\maketitle
 \author {Francesco Paparella,$^{1\ast}$ Marcello Vichi$^{2}$\\ \\ \normalsize{$^{1}$Division of Sciences and Mathematics, Center on Stability, Instability and Turbulence, New York University Abu Dhabi,}\\ \normalsize{Saadyiat Island, Abu Dhabi, UAE}\\ \normalsize{$^{2}$Department of Oceanography and Marine Research Institute, University of Cape Town}\\ \normalsize{Cape Town, Rondebosch 7701, South Africa}\\ \\ \normalsize{$^\ast$To whom correspondence should be addressed; E-mail:  francesco.paparella@nyu.edu.} }\pagebreak
\begin{abstract}
The quantitative description of marine systems is constrained by a
major issue of scale separation: most marine biochemical processes
occur at sub-centimeter scales, while the contribution to the Earth's
biogeochemical cycles is expressed at much larger scales, up to the
planetary one. In spite of vastly improved computing power and observational
capabilities, the modeling approach has remained anchored to an old
view that sees the microscales as unable to substantially affect larger
ones. The lack of a widespread theoretical appreciation of the interactions
between vastly different scales has led to the proliferation of numerical
models with uncertain predictive capabilities. We show that an enhanced
Lagrangian modeling framework, allowing for those interactions, can
easily tackle puzzling problems such as the phenology of phytoplankton
blooms, or vertical variability in mixed layers. 
\end{abstract}

\section*{Introduction}

Marine phytoplankton is involved in several biogeochemical processess
at the microbial ocean scale that affect entire ecosystems \cite{azam2004,legendre2018}.
Predictive models of phytoplanktonic processes are thus fundamental
to many applications. In climate projections, the ``biological pump''
is a fundamental component of the carbon cycle \cite{gruber2009,bopp2013},
described by modeling phytoplankton primary production and the net
export of organic matter through the marine food web and the water
column. Models of biogeochemical and phytoplankton processes are also
employed in operational oceanography and coastal management \cite{piroddi2015,hyder2015}.
Predictive models of coastal and near-shore transport are coupled
with water quality and biogeochemical models to provide forecasts
of undesirable disturbances such as eutrophication, hypoxia, or harmful
algal blooms. Ultimately, the outputs of these models are used for
fishery management, end-to-end ecosystem models, and indicators of
ocean health \cite{travers2007,fu2018}.

There is, however, a fundamental difficulty in the modeling process:
namely, the chasm between the scales where the biogeochemical processes
occur and are being observed (by probes deployed in the ocean, laboratory
experiments or metagenomics studies \cite{azam2004,stec2017,legendre2018})
and the scales where the system response is sought, observed and interpreted
(by remote sensing, data aggregation and models). Laboratory experiments
such as cultures and mesocosms allow to empirically estimate a model's
biological terms, upon the assumption of homogeneous distribution
of all the biochemical fields \cite{denman2003,tian2006}, while neglecting
the physical terms. The interactions between these terms is ultimately
mandated to the numerical solution of coupled physical-biogeochemical
models \cite{nihoul1975,nihoul1998}, which cannot include all the
spatial and temporal scales necessary to close the chasm.

In this paper, we address the theory behind marine physical-biogeochemical
models, we expose some limits of the current models, and we propose
a new approach. To make our point, we focus on the open ocean mixed
layer and phytoplankton dynamics, which is at the base of the water
column biogeochemistry \cite{legendre2018}.

The distribution of plankton shows variability from the global scale
down to the microscale (centimetric lengths) \cite{pinel-alloul2007,prairie2012}.
Plankton patchiness at the mesoscale and sub-mesoscale is shaped by
the interaction between biological growth processes and turbulent
lateral stirring linked to upper ocean frontal eddies and currents
\cite{martin2003,mahadevan2016,levy2018}. Lateral stirring and mixing
alone cannot generate patchiness \cite{martin2003}. A triggering
mechanism is needed, and physics-driven processes affecting the vertical
structure of the mixed layer (that is, on scales smaller than 100
m) may easily fulfill this role. This variability, in turn, is enhanced
by biological processes such as the interplay between light and nutrient
gradients, cell buoyancy adjustments, gyrotaxis, convergent swimming,
and light-dependent grazing \cite{huisman2006,durham2011,cullen2015,moeller2019}.
The emerging very-high-resolution sampling techniques suggest that
plankton remains patchy at the scales of one meter both in the vertical
and in the horizontal \cite{foloni-neto2016}, and that homogeneity
might not be reached before the centimeter scales \cite{currie2006,doubell2009,foloni-neto2016}.
As we shall see, models assuming homogeneity at the fine and micro
scales may easily incur in serious biases.

Three classes of processes should be included to model marine biogeochemical
processes at the microscale: turbulent stirring, caused by fluid eddies,
which displaces, stretches, and folds water volumes, increasing the
gradients of the transported fields; irreversible mixing, caused by
sub-microscale processes, which decreases these gradients; and growth
(or decay) which changes the concentration of the fields by chemical
or biological means. In principle, these processes must correspond
to distinct terms in the equations resolved by numerical models. In
practice, two broadly-defined formulation may be used to build a model:
the Eulerian and the Lagrangian. Choosing the formulation binds the
model equations to conform to a set of approximations which may not
strictly abide to this principle.

\begin{figure}
\begin{centering}
\includegraphics[width=1\textwidth]{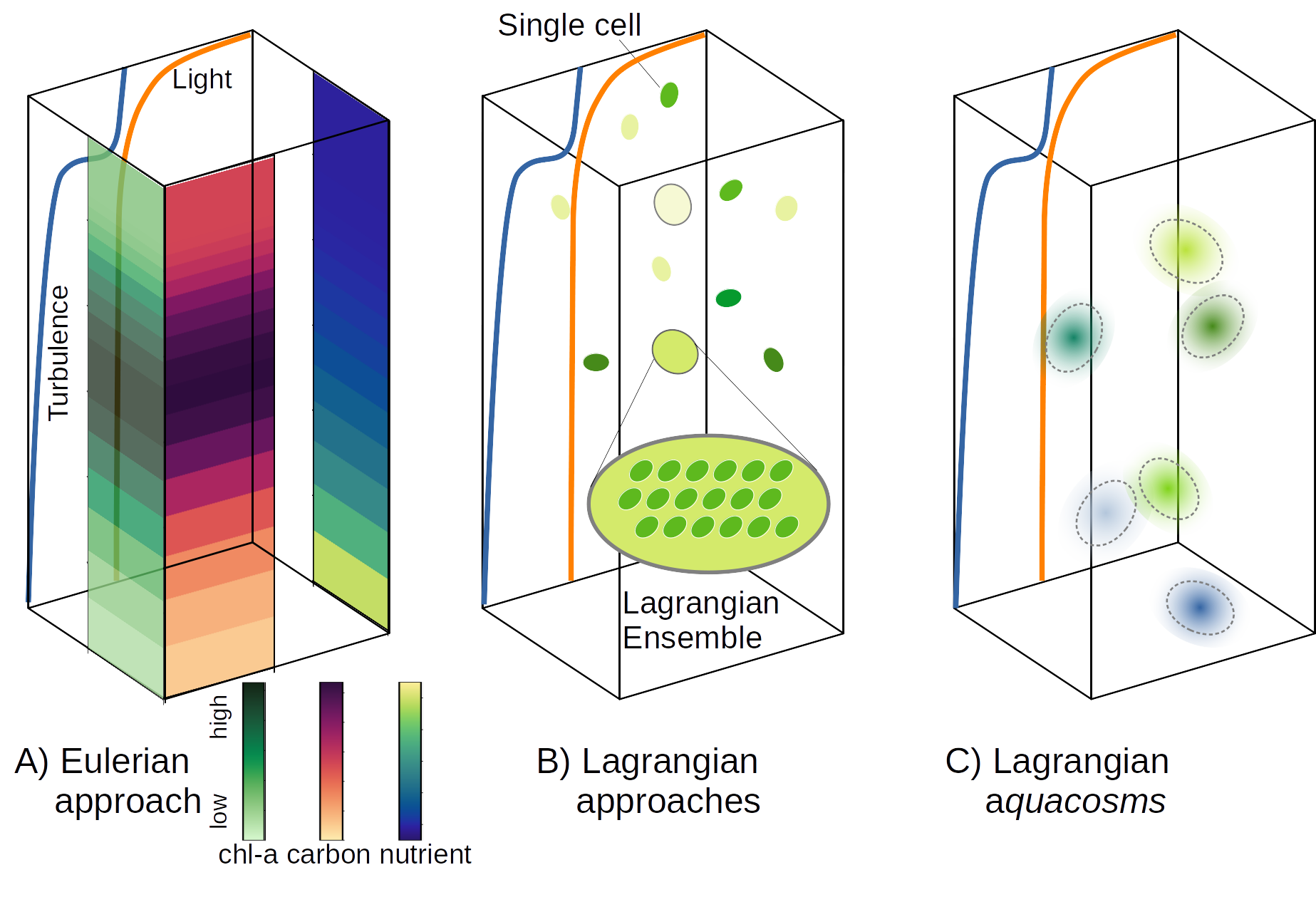}
\par\end{centering}
\caption{\label{fig:Schematic}Schematic of different phytoplankton modeling
approaches. \textbf{A)} The vertical distribution of phytoplankton
carbon and chlorophyll can be simulated according to vertical nutrient
and light gradients and a turbulent field. The Eulerian approach samples
all active fields at the nodes of a fixed grid; fine and microscale
stirring (turbulence) is modeled as irreversible mixing, wiping out
fine and microscale patchiness. \textbf{B)} The Lagrangian Ensemble
approach bundles individual cells into Lagrangian particles; the number
of organisms per particle is modified by infra-particle ecological
interactions; unresolved turbulence is modeled as a stochastic motion
of the particles, which don't interact with each other. \textbf{C)}
The Lagrangian aquacosm approach tracks tiny Lagrangian water masses
(\emph{aquacosms}) moving as in B); biogeochemical interactions occur
within aquacosms, which are permeable, thus allowing for mass exchanges
between nearby aquacosms.}
\end{figure}

The majority of model applications mentioned above are Eulerian \cite{denman2003,lequere2005,vichi2007,aumont2015}.
In this formulation, all three processes occur at the nodes of a fixed
spatial grid, where all the relevant fields are located (Fig. \ref{fig:Schematic}A).
Invoking the continuum approximation, the biological variables are
approximated as smoothly varying mean fields whose values at the grid
nodes are representative of the average values in the grid cell \cite{vichi2007}.
An important feature of Eulerian models is that the unresolved turbulent
stirring processes are assimilated to irreversible mixing. While this
practice may achieve satisfactory results for non-reacting, passively
transported tracers, it yields questionable, if not flawed, results
for biological and chemical tracers, because it hopes that the biological
response to the simulated Eulerian mean field is the same as the average
response to the real, unresolved, patchy environment \cite{paparella2018,baudry2018}.

Other models use the Lagrangian formulation, which singles-out either
small portions of the fluid or individual biological agents, and follows
them along their motion (Fig. \ref{fig:Schematic}B). Despite their
approximations (e.g. number of particles insufficient to resolve all
the fluid structures and use of stochastic processes to mimic turbulence),
Lagrangian models describe stirring processes as such, rather than
assimilating their effect to irreversible mixing. In plankton modeling,
the Lagrangian formulation, originally identified with the term Lagrangian
ensemble \cite{woods1982,wolf1988,woods1994,woods2005}, is often
referred to as individual based modeling \cite{cianelli2012}. We
argue that a clear distinction should be made between single cell
Lagrangian models \cite{yamazaki1991,kamykowski1994}, in which the
movement of a single individual is followed, but growth/death processes
and cell division are not included, and Lagrangian ensembles (Fig.
1B), where the super-individual concept \cite{scheffer1995} is used
to describe the plankton population dynamics. Some authors have stated
the superiority of the Lagrangian approach in describing plankton
dynamics \cite{woods2005,hellweger2007,hellweger2009a,baudry2018}.

However, there are intrinsic limitations to applying the super-individual
concept to the modeling of phytoplankton communities, because a Lagrangian
ensemble is not conceived to exchange with surrounding ensembles any
of the active agents that it carries (Fig. \ref{fig:Schematic}B).
Therefore, nearly all Lagrangian models (but see \cite{dippner1998})
neglect to include irreversible mixing processes. In the sporadic
cases where Lagrangian and Eulerian formulations have been compared,
this issue appears to have been overlooked \cite{wolf1988,lande1989,mcgillicuddy1995,kida2017lagrangian,baudry2018},
even though it may lead to unrealistic, even paradoxical outcomes.

Consider a region of ocean with steady conditions, favorable for a
phytoplankton bloom. Assume an initial random distribution dividing
the fluid in very small patches, half devoid of phytoplankton, and
the others at carrying capacity. The ensembles of a Lagrangian model
would mimic these patches, but, lacking any mutual interaction, plankton
in those already at carrying capacity would never reach the nearby
empty ones and trigger growth. The ensembles lacking phytoplankton
would remain devoid of it, and the others would stay at the carrying
capacity. The bulk concentration, computed as an average over all
the Lagrangian ensembles, would indefinitely remain at one half of
the carrying capacity: a baffling outcome given the favorable conditions!
In an Eulerian model, irreversible mixing would quickly offset from
zero the concentration of the empty nodes, triggering growth, so that
the bulk concentration will eventually reach the carrying capacity.
However, if the initial subdivision of empty and full patches were
too fine to be resolved, then the amount of irreversible mixing computed
by the Eulerian model would be a gross overestimation of the real
one, which begs the question whether the modeled growth rate of the
bulk concentration is realistic \cite{baudry2018}. 

Turbulent stirring, irreversible mixing, and growth are each associated
to their own distinct time scales. For example, the celebrated Sverdrup
model \cite{sverdrup1953} for the onset of phytoplanktonic blooms
stems from the assumption that the growth time scale is slower than
the stirring time scale. It is of extreme historical importance, and
is the founding stone that all later bloom models have confronted,
either to build on it, or to overthrow it \cite{fischer2014,sathyendranath2015}.
It also has a peculiar feature: owing to its linearity, substituting
stirring with irreversible mixing (if characterized by the same time
scales as the stirring) leaves the results unchanged. As we shall
illustrate in the following, in the presence of nonlinear biological
terms, this equivalence is lost: in Sverdrup-like nonlinear models,
the separation between the time scales of stirring and of irreversible
mixing determines the tempo and mode of the bulk phytoplankton growth.
The proliferation of explanations for the occurrence of blooms, often
distinct from each other by subtle details, may be a symptom of the
lack of appreciation for a key theoretical issue: phytoplankton patchiness
affects the bulk growth. 

Occasionally, some attempts have been made to parameterize patchiness
effects into Eulerian biogeochemical models. Realizing that the biological
response is greatly affected by the treatment of the unresolved scales,
authors like Fennel \cite{fennel1996} have long proposed to use ``effective''
biological parameters. A time–delay parameterization was suggested
for the case where patchiness is the result of an oscillatory population
dynamics occurring with different phases in different places \cite{wallhead2006}.
More recently, a closure parameterization was introduced, reminiscent
of those used for turbulence \cite{mandal2016}. The approach is formally
valid only when the fluctuations are small, which high resolution
chlorophyll profiles suggest is not the case \cite{doubell2014}.
Overall, the bulk of the literature appears to overlook the issue,
treating biogeochemical tracers in the same way as non reacting ones.

We argue that the strategy of replacing unresolved transport with
irreversible mixing, and then compensate the resulting biases by means
of some parameterization, will face overwhelming difficulties. For
example because different initial conditions, keeping everything else
the same, may yield different bulk growth rates \cite{paparella2018}
(an issue also noted in the early work on the plankton patchiness
theory \cite{martin2003}).

If turbulent stirring, irreversible mixing, and growth processes are
modeled separately and independently from each other, then reproducing
realistic phytoplankton dynamics in predictive models should become
much easier. A class of Lagrangian methods recently proposed \cite{paparella2018}
achieves this goal by depicting Lagrangian particles as representing
microscale-sized, homogeneous control volumes of water, rather than
individual organisms or ensembles. In this framework, irreversible
mixing processes are represented by exchanging small mass fluxes between
nearby particles. We call \emph{aquacosms} such Lagrangian particles
subject to coupling fluxes (Fig. \ref{fig:Schematic}C, see ``methods''
for details). The coupling is regulated by a parameter, $p$, whose
value is proportional to the intensity of the fluxes. As we shall
demonstrate, $p$ sets the time scale associated with the irreversible
destruction of biogeochemical variance at the microscales, which,
in this approach, remains independent of the time scales of mechanical
stirring. Results analogous to those of Lagrangian ensemble models
are recovered for uncoupled particles ($p=0$). In the opposite limit,
high values of $p$ produce an excessive irreversible mixing, and
yield results strongly resembling Eulerian simulations.

\section*{Results}

In the following we examine the results obtained with water-column
models, considering first a few idealized cases, and then more realistic
open ocean situations from the North Pacific and the Southern Ocean.
We compare an Eulerian model, where vertical turbulent stirring is
parameterized by eddy diffusivity, with Lagrangian aquacosm models,
distinguished by different intensities of the microscale irreversible
mixing, where vertical turbulent stirring is modeled as stochastic
motions of the aquacosms, matching the Eulerian eddy diffusivity.

\subsection*{Pure stirring and mixing}

\begin{figure}
\begin{centering}
\includegraphics[width=1\textwidth]{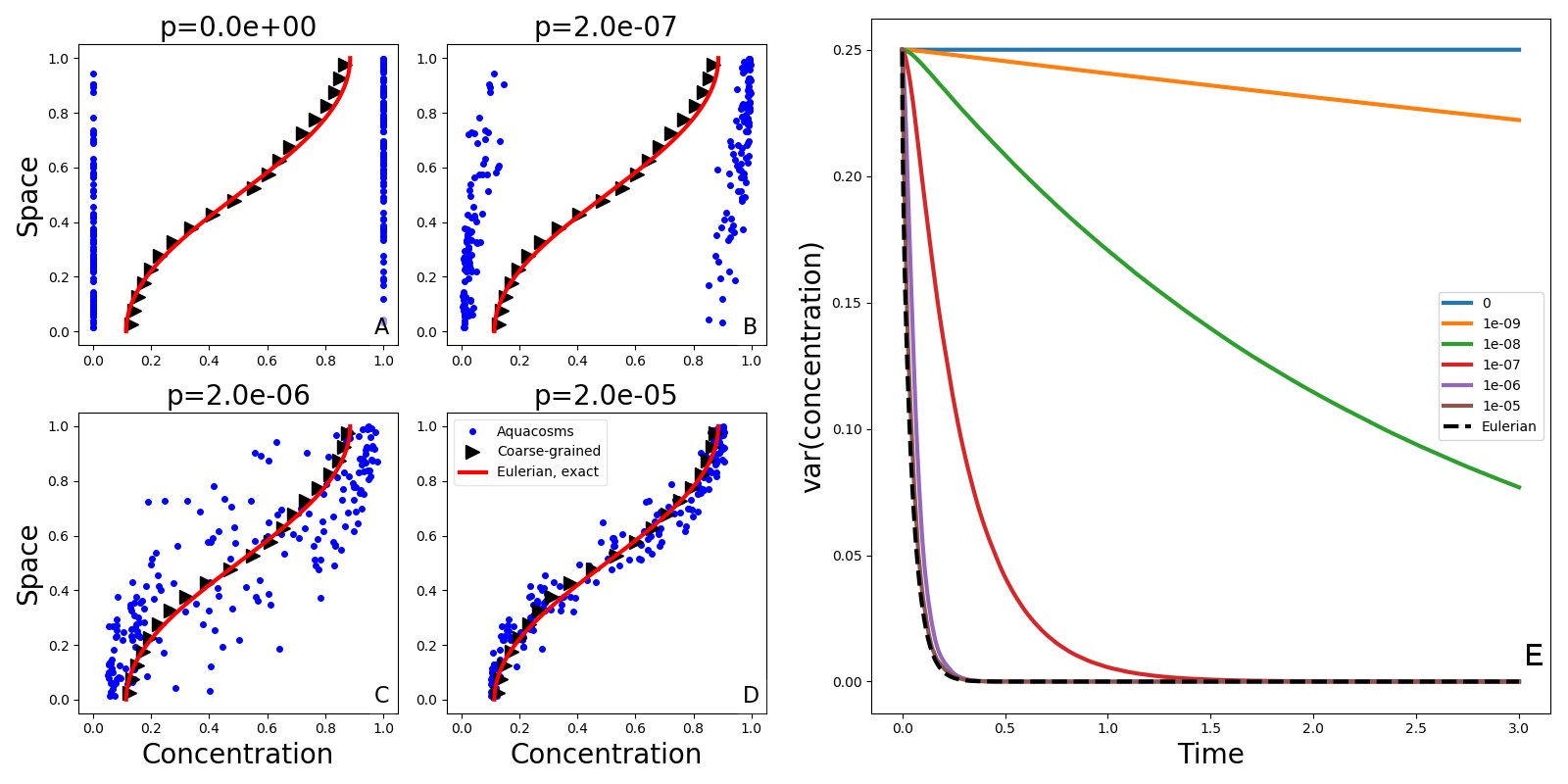}
\par\end{centering}
\caption{\label{fig:stirring_vs_mixing}\textbf{A-D)} Exact solutions at time
$t=0.05$ of the diffusion equation (\ref{eq:diffusion_equation})
with the initial condition (\ref{eq:step_ic}) and no-flux boundary
conditions (red lines). The blue dots show the position and the concentration
of the Lagrangian particles for equivalent aquacosm simulations with
the same diffusivity and varying coupling strength $p$. The black
triangles are a coarse-grained version of the same data, obtained
with a Gaussian kernel estimator with standard deviation 0.1 (see
methods). \textbf{E)} variance, as a function of time, in Lagrangian
aquacosm simulations with varying coupling strength $p$ (solid lines)
and variance of the exact solution of the Eulerian problem (\ref{eq:diffusion_equation})
(dashed line).}
\end{figure}

First, we shall consider the case of turbulent stirring and mixing
of a substance not subject to any reaction. In non-dimensional units,
the eddy-diffusive Eulerian model is 
\begin{equation}
\frac{\partial C}{\partial t}=\frac{\partial^{2}C}{\partial z^{2}}\label{eq:diffusion_equation}
\end{equation}
with $C$ representing the concentration of the non-reactive substance,
subject to no-flux boundary conditions in the water column $0<z<1$.
In this formulation, turbulent stirring is completely replaced by
irreversible mixing, described mathematically by the right-hand side
term of the equation. As the initial condition, we choose the step
function 
\begin{equation}
C(z,0)=\begin{cases}
0, & z<\nicefrac{1}{2}\\
1, & z\ge\nicefrac{1}{2}
\end{cases}.\label{eq:step_ic}
\end{equation}

In the Lagrangian aquacosm approach, just as in other Lagrangian methods,
turbulent stirring is modeled as a Brownian motion which scrambles
the position of the particles. Irreversible mixing is modeled separately
and its intensity is determined by the value of the coupling parameter
$p$ (see methods). The aquacosms (Lagrangian particles) are initially
placed at uniformly random position within the domain, those in the
first half take a concentration of zero, the others take a concentration
of one.

Figures \ref{fig:stirring_vs_mixing}A-D show the concentration and
position of the aquacosms for different values of the coupling parameter
$p$, together with the analytical solution of equation (\ref{eq:diffusion_equation}).
Brownian motion produces patchiness: a random alternation of particles
with high and low concentration values. Irreversible mixing equalizes
the concentration of nearby particles, thus removing patchiness, more
and more effectively as the coupling strength $p$ increases. Local
weighted averages of the Lagrangian result (black triangles, see methods),
are essentially identical to the analytic solution of the Eulerian
model. This coincidence might suggest that the Eulerian and all of
the Lagrangian models are equivalent. The coarse-graining process
of taking local averages, however, gives only a partial picture. Panels
A-D do not depict equivalent microphysics. The concentration carried
by the individual particles (which, ultimately, is all that matters
for the reaction terms when they are present) is vastly different
in the four cases. The amount of irreversible mixing, set by the parameter
$p$, determines how quickly the fluctuations around the local averages
are dissipated (Fig. \ref{fig:stirring_vs_mixing}E), that is, $p$
sets the time scale associated to irreversible mixing processes. When
irreversible mixing is strong (high values of $p$) the fluctuations
around the local means, thus variance, decay just as quickly as in
the Eulerian case. For smaller values of $p$ the time scale of variance
decay can be made arbitrarily large: fluctuations may persist even
when the local averages suggest the presence of a vertically near-uniform
profile of concentration. In the limit case $p=0$ (uncoupled particles)
each aquacosm maintains its initial concentration, and variance remains
constant in time, even though the local averages still tend to uniformity,
as prescribed by (\ref{eq:diffusion_equation}).

\subsection*{Sverdrup's model expanded}

Next we add to equation (\ref{eq:diffusion_equation}) a simple logistic
growth term. In dimensional units our model is 
\begin{equation}
\frac{\partial C}{\partial t}=\kappa\frac{\partial^{2}C}{\partial z^{2}}+rf(z)C\left(1-\frac{C}{K}\right)\label{eq:Logistic_dimensional}
\end{equation}
where $\kappa$ is the eddy diffusivity, assumed to be constant, and
$r$ is the maximum growth rate of phytoplankton, having concentration
$C$ and carrying capacity $K$. Zero-flux boundary conditions are
imposed at both ends of the water column, that is at $z=0$ and $z=\ell$.
This model is more general than Sverdrup's original one, as it does
not yet make any assumption on the relative size of the turbulent
stirring and biological time scales, and it includes a nonlinear growth/loss
term. However, it still parameterizes stirring with an eddy diffusion
term, thus interchanging stirring with irreversible mixing. The non-dimensional
function $f$ quantifies the balance between light-stimulated growth,
and loss of biomass due to respiration. Sverdrup defined it as 
\begin{equation}
f(z)=e^{-\lambda z}-\frac{\mu}{r},\label{eq:Sverdrup_f}
\end{equation}
where $\mu$ is a constant respiration rate, and $\lambda$ is a measure
of water transparency, but Sverdrup's argument holds for any integrable
function $f$ sandwiched between $O(1)$ bounds. Choosing $\ell$
as the unit of length, $\ell^{2}/\kappa$ as the unit of time, and
$K$ as the unit of concentration, equation (\ref{eq:Logistic_dimensional})
takes the non-dimensional form: 
\begin{equation}
\frac{\partial C}{\partial t}=\varepsilon f(z)C(1-C)+\frac{\partial^{2}C}{\partial z^{2}}.\label{eq:Logistic_non-dimensional}
\end{equation}
The parameter $\varepsilon=r\ell^{2}/\kappa$ expresses the ratio
of the stirring/mixing and the biological time scales (the former
quantified through the eddy diffusivity as $\ell^{2}/\kappa$). Formally,
Sverdrup's approximation applies for vanishing $\varepsilon$, when
one can argue that physics and biology disentangle: in that case,
the eddy diffusion term makes the initial condition vertically homogeneous
after a transient no longer than $O(1)$, keeping the concentration
$C$ independent of depth at all later times. Then, on $O(\varepsilon^{-1})$
time scales, the vertically constant concentration changes in time
according to the ODE
\begin{equation}
\dot{C}=\varepsilon I\,C(1-C)\label{eq:ODE}
\end{equation}
where $I$ is the integral of $f$ over the water column and the dot
denotes a derivative with respect to time. In practice, this argument
yields a good approximation up to $\varepsilon\approx1$.

In our Lagrangian formulation of this problem the aquacosms move by
performing a Brownian motion. The concentration in each aquacosm is
determined by the ODE associated with the reaction term in (\ref{eq:Logistic_non-dimensional}),
and by the coupling fluxes modeling irreversible mixing (see methods).
If the growth term is linearized, then the eddy-diffusive Eulerian
and the Lagrangian approach are equivalent in a coarse-grained sense
(as partly observed by other authors \cite{lande1989,mcgillicuddy1995}),
but we show here, and demonstrate analytically in sec. S1 of the supplementary
materials, that nonlinear biological terms break down the equivalence.
\begin{figure}
\begin{centering}
\includegraphics[width=1\textwidth]{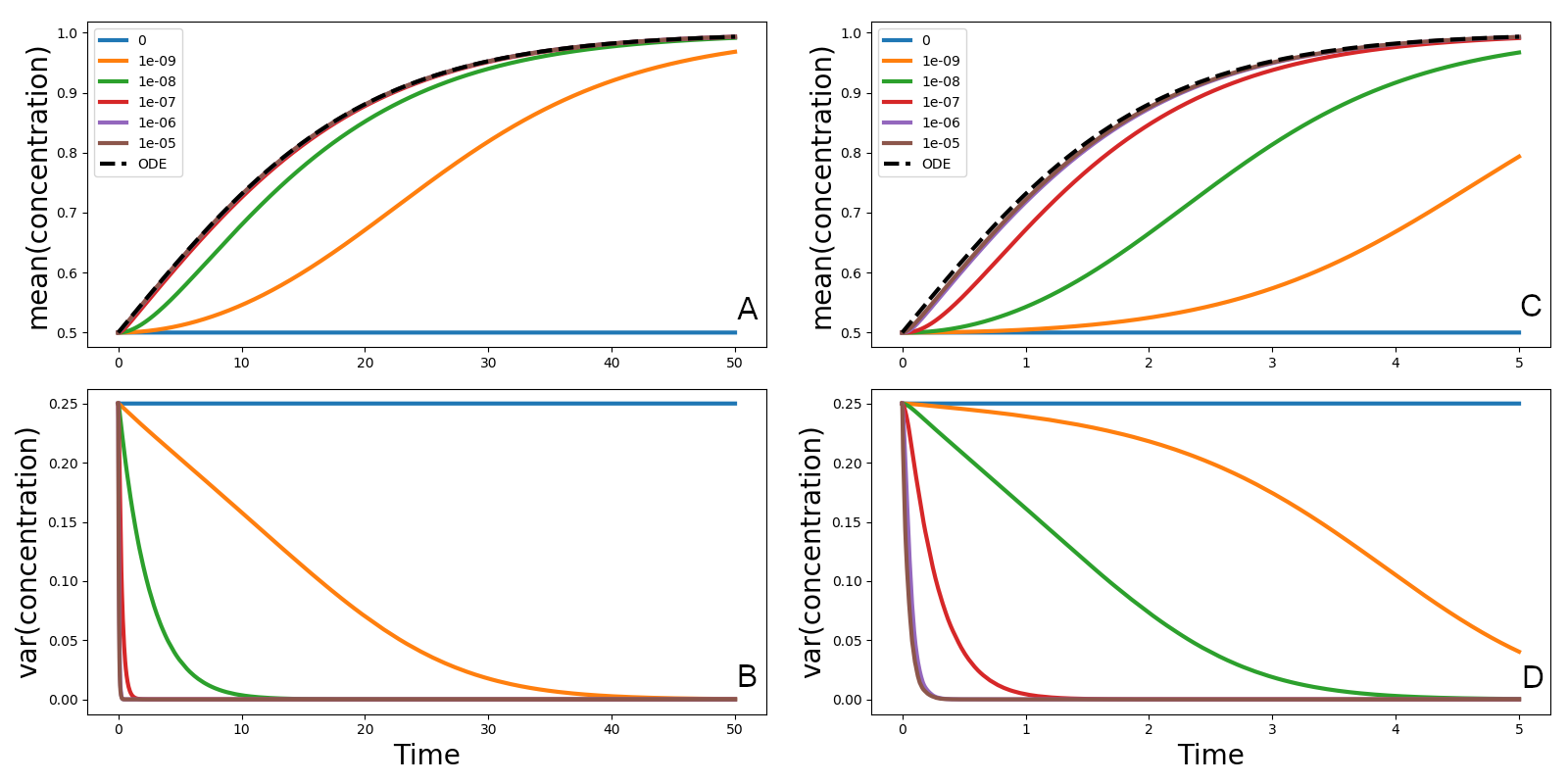}
\par\end{centering}
\caption{\label{fig:Logistic}\textbf{(A, C)} Average plankton concentration
and \textbf{(B, D)} variance, as a function of time, for the problem
(\ref{eq:Logistic_non-dimensional}) with $f(z)=1$, $\varepsilon=0.1$
(A, B) or $\varepsilon=1$ (C, D) and the step initial condition (\ref{eq:step_ic}).
The solid lines refer to Lagrangian aquacosm simulations with varying
strength of the coupling parameter $p$. The dashed line is the solution
of Sverdrup's approximation (\ref{eq:ODE}). Note the different ranges
of the time axes.}
\end{figure}

Figure \ref{fig:Logistic} shows the phytoplankton concentration mean
and variance as a function of time for equation (\ref{eq:Logistic_non-dimensional})
and its aquacosm counterpart, with $f(z)=1$, the initial condition
(\ref{eq:step_ic}), and two values of the ratio $\varepsilon=0.1$
and $\varepsilon=1$. As sketched in the introduction, because all
fluid parcels are initially set at a fixed point of the reaction terms,
with uncoupled particles ($p=0$, as in a Lagrangian ensemble model)
the mean unrealistically remains at one half of the carrying capacity
despite the positive growth rate. When the coupling between the aquacosms
is switched on, the mean gradually tends to the carrying capacity
($C=1$) and the variance tends to zero. The rapidity with which the
asymptotic values are approached depends on the strength of the coupling
parameter $p$: high values produce results that behave just as predicted
by Sverdrup's theory, but small ones produce growth curves that are
nothing like the solution of the ODE associated with the reaction
term. Because Sverdrup's theory postulates the equivalence of stirring
and irreversible mixing, the time scales of the two kind of processes
are the same: a water parcel devoid of plankton and one full of plankton
equalize their concentration within the stirring time scale. When
stirring and mixing are treated as separate processes, with the second
allowed to be slower than the first, parcels lacking plankton are
constantly seeded by the full ones, and the overall rate of growth
is determined by a delicate interplay of biological processes, turbulent
stirring and irreversible mixing. Lagrangian ensemble models only
account for the first, and eddy-diffusive Eulerian models are dominated
by the second. We notice that in a generic non-homogeneous environment
with concurrent stirring, mixing and growing, there is no reason to
expect that the bulk phytoplankton concentration is described solely
by the ODE associated with the reaction terms.

\begin{figure}
\begin{centering}
\includegraphics[width=1\columnwidth]{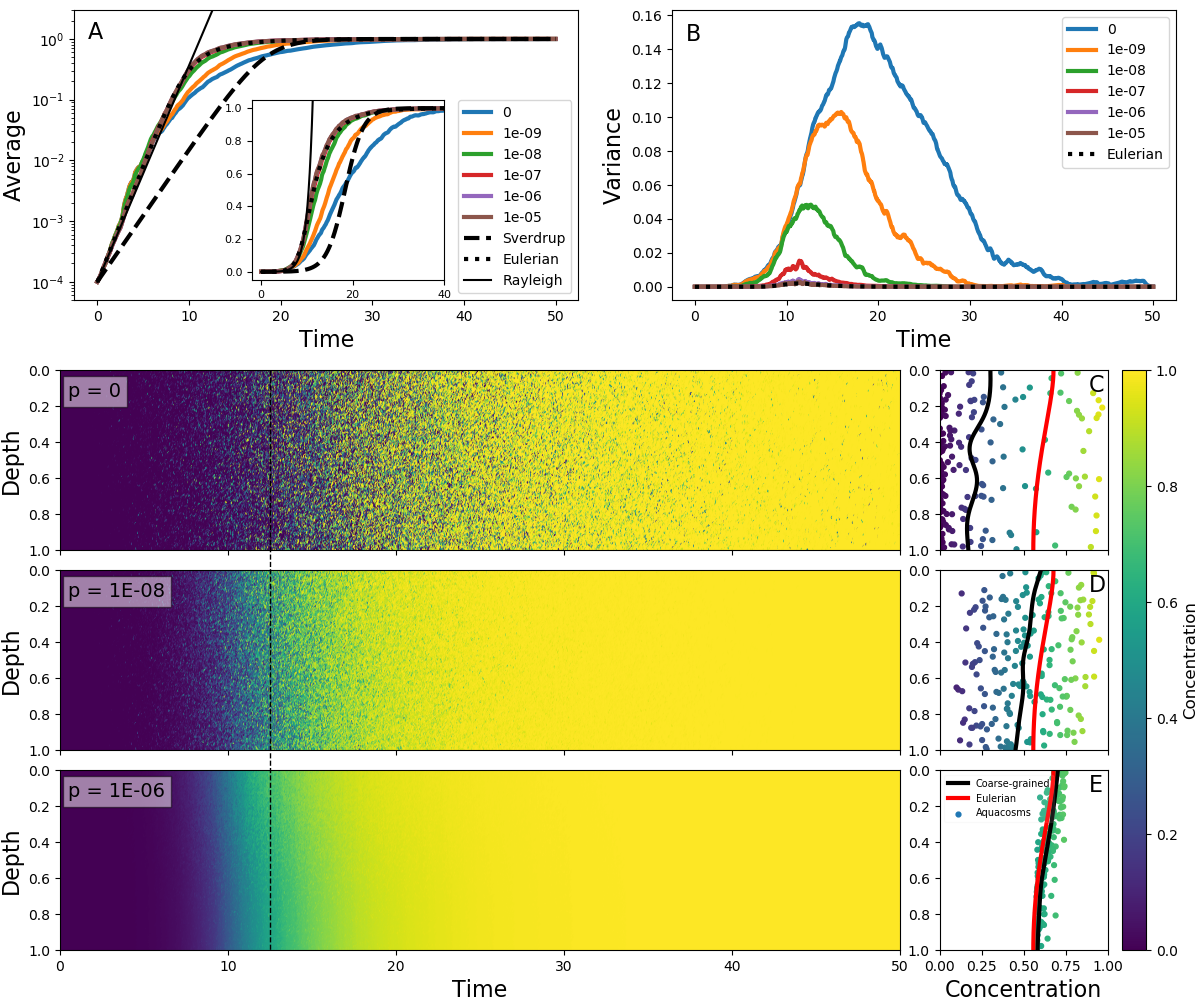}
\par\end{centering}
\caption{\label{fig:Logistic_Sverdrup}\textbf{(A)} Average phytoplankton concentration
and \textbf{(B)} variance as a function of time, for the Eulerian
problem (\ref{eq:Logistic_non-dimensional}) with $f(z)=e^{-\nicefrac{z}{0.15}}-0.1$,
$\varepsilon=10$ and constant-in-space initial condition $C(z)=10^{-4}$
(dotted lines). The solid lines in color refer to Lagrangian aquacosm
simulations with the same parameters, with varying coupling strength
$p$. The dashed line is the solution of Sverdrup's approximation
(\ref{eq:ODE}). The thin black line is an exponential growth with
a rate estimated by means of the Rayleigh quotient (sec.S2 in supp.
mat.) associated to equation (\ref{eq:Logistic_non-dimensional}).
The inset shows the same data as panel A, but with a linear, rather
than logarithmic, vertical axis.\textbf{ (Lower panels)} plankton
concentration as a function of depth and time for Lagrangian simulations
with $p=0,10^{-8},10^{-6}$. \textbf{(C-E)} concentration and depth
of the aquacosms (dots), coarse-grained concentration of the aquacosms
(black line) and numerical solution of the Eulerian model (red line)
at the time marked by the dashed black line in the left panels.}
\end{figure}

The examples so far show that the fast stirring of a step-like initial
condition produces patchiness, which then affects the bulk growth
rate. Much attention has been given to the case of non-homogeneous
biological growth (e.g. using the form (\ref{eq:Sverdrup_f}) for
$f\left(z\right)$) when growth is faster than stirring. Then, according
to the critical turbulence theory \cite{huisman1999critical,taylor2011shutdown,ferrari2015shutdown},
phytoplankton closer to the surface contributes to the overall growth
more than Sverdrup's theory would provide for, so that a bloom may
initiate even when the average light would not allow for that. Differences
in light history and acclimation have also been affirmed to produce
growth when Sverdrup's theory would predict decay \cite{lande1989,woods1994,esposito2009}.
What has not been stressed is that these conditions would also naturally
lead to the creation of patchiness if irreversible mixing processes
are not fast enough to remove it. When growth is faster than turbulent
stirring, then the phytoplankton in water parcels at shallow depth
will have time to grow substantially more than that in the deeper
parcels spending some time in darkness. As stirring makes some of
the shallow particles sink and replaces them with some of those that
were at depth, then microscale patchiness ensues. Just as in the case
of a step-like initial condition, patchiness affects growth. As the
bloom progresses, water parcels having spent the most time close to
the surface reach the carrying capacity before the others, and the
rapidity of the bulk growth becomes regulated by the intensity of
the irreversible mixing, which transfers plankton from high to low
concentration particles.

This process is illustrated in Fig. \ref{fig:Logistic_Sverdrup} for
different degrees of the coupling between aquacosms. In the linear
regime, when phytoplankton concentration is much smaller than the
carrying capacity, the Eulerian model (\ref{eq:Logistic_non-dimensional})
and all the Lagrangian models show the same growth rate of the bulk
concentration (Fig. \ref{fig:Logistic_Sverdrup}A), which is in excess
of what Sverdrup's theory would dictate. This is in agreement with
the critical turbulence prescriptions (see sec. S2 of supplementary
materials) as long as the process is linear. In the nonlinear regime,
the Lagrangian models yield distinct results: with small coupling
strength, variance grows with time (Fig. \ref{fig:Logistic_Sverdrup}B),
and this leads to bulk growth significantly slower than in the eddy-diffusive
Eulerian model, due to the patchy environment. Once again, destroying
the variance by using a strong enough coupling parameter recovers
the results of the Eulerian model. Figure \ref{fig:Logistic_Sverdrup}C-E
shows that variance genuinely corresponds to patchiness: with low
$p$, aquacosms of starkly different concentration are found next
to each other, and the coarse-grained equivalence of Lagrangian and
Eulerian models is lost.

\subsection*{Microscales and phytoplankton phenology}

To demonstrate the effect of irreversible mixing on phytoplankton
phenology, we run an eddy-diffusive Eulerian model and Lagrangian
aquacosm models. They use realistic eddy diffusivity profiles and
incident solar radiation data spanning one year, representative of
the conditions found at Ocean Station PAPA in the north-east Pacific
Ocean and in the sub-Antarctic zone of the Southern Ocean, and a simplified
version of a biogeochemical model currently used in ocean climate
simulations, where light availability is the only explicit limiting
factor, and a crowding mortality term parameterizes zooplankton grazing
(see methods).

\begin{figure}
\begin{centering}
\includegraphics[width=1\columnwidth]{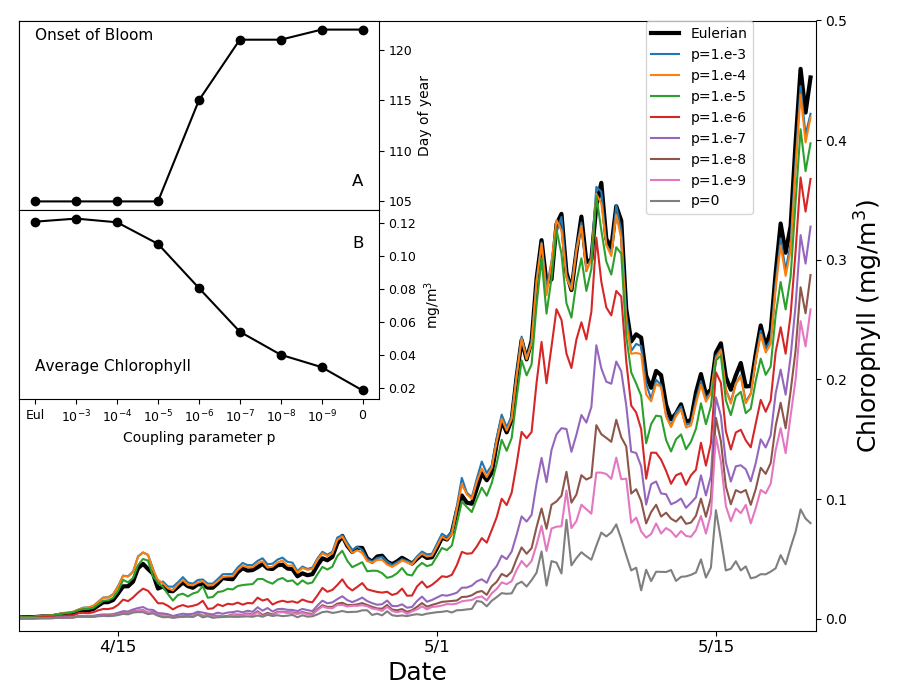}
\par\end{centering}
\caption{\label{fig:PAPA_phenology}Average chlorophyll in the first 50 m of
water column for the simulated Ocean Station PAPA in the North-West
Pacific. Thin colored lines refer to aquacosm simulations coupled
with varying strength $p$. The thick back line refers to the eddy-diffusive
Eulerian model. Inset A shows the day of the onset of the bloom (see
main text); inset B shows the average chlorophyll in the period April
15th to May 15th, both of them plotted as functions of the coupling
parameter.}
\end{figure}
Figure \ref{fig:PAPA_phenology} shows the average chlorophyll content
at PAPA station over the first 50 m of water column in the weeks when
the bloom starts. We define the onset of the bloom as the first day
of the year when the chlorophyll content exceeds the median plus 5\%
of the daily chlorophyll concentration tracked over one year \cite{racault2012}.
Reductions of the coupling strength $p$ delay the onset by over two
weeks. The results diverge from each other before the onset of stratification,
from mid April until the beginning of May, when the mixed layer depth
is $\ell\approx100$ m (supplementary Fig. \ref{fig:PAPA_chlorophyll})
and the typical mixed layer eddy diffusivity is $\kappa\approx0.05$
m$^{2}$s$^{-1}$, yielding a value of $\varepsilon\approx5$ with
the growth rate $r=2$ days$^{-1}$. This suggests that we are witnessing
the same process illustrated in Figure \ref{fig:Logistic_Sverdrup},
where low irreversible mixing allows the formation of high microscale
patchiness, reducing bulk growth and delaying the bloom. The Lagrangian
model without any coupling shows the slower growth, but this approach
is unrealistic, as demonstrated above. In the following 15 days the
mixed layer becomes much shallower (supplementary Fig. \ref{fig:PAPA_chlorophyll}),
with typical eddy diffusivity values of $\kappa\approx0.025$ m$^{2}$s$^{-1}$.
The time scales of growth and stirring become comparable, and the
vertical light gradient ceases to be a source of patchiness. Only
the entrainment of phytoplankton-poor aquacosms at the base of the
mixed layer continues to be a source of patchiness. Overall, the chlorophyll
content averaged over the month of the bloom initiation changes by
as much as a factor 6 depending on the coupling strengths (inset B
in Fig. \ref{fig:PAPA_phenology}).

Next, we simulate the open ocean of the sub-Antarctic zone (SAZ),
characterized by a mixed layer deeper than 100 m from July to October
(Fig. \ref{fig:SAZ_phenology}). The biological model is the same
as for the PAPA simulations, but, owing to the colder water temperature,
different parameter values are used, in particular the maximum photosynthetic
rate is set to $r=0.5$ days$^{-1}$ (see methods). Throughout the
year, typical simulated eddy diffusivity values in the mixed layer
are $\kappa\approx0.06$ m$^{2}$s$^{-1}$, corresponding to $\varepsilon\approx1$.
Here, growth is never faster than stirring, and Sverdrup's approximation
holds. Therefore, only the intermittent deepening of the mixed layer,
which scoops phytoplankton-free aquacosms from the depths, contributes
to the creation of patchiness in the mixed layer. In the days immediately
after a sudden deepening of the mixed layer, the dynamics is reminiscent
of that shown in Fig. \ref{fig:stirring_vs_mixing} and \ref{fig:Logistic},
whereby a step-like initial condition first breaks down into patchiness
and then is brought to vertical homogeneity with a speed determined
by the intensity of the irreversible mixing. The smaller is the mixing,
the slower is the destruction of variance, and the slower is the bulk
growth of phytoplankton (supplementary Fig. \ref{fig:SAZ_phenology}).
The phenological and productivity differences are not as marked as
in the PAPA simulations, but we note that coarser resolution climate
models with a larger impact of irreversible mixing are likely to generate
greater differences and discrepancies in the bloom phenology \cite{hague2018}.

Recent bio-optical measurements using ARGO floats from sub Antarctic
zones \cite{carranza2018} reported the presence of substantial chlorophyll
variance within the hydrographic mixed layer. This was interpreted
as the signature of vertical gradients of chlorophyll at the fine
scales (tens of meters), which called for some mechanism incompatible
with the presence of strong turbulence. In particular, it was argued
that periods of storm quiescence associated with slacking turbulence
would occasionally leave the mixed layer homogeneous in density, but
stirred only in its uppermost part, thus allowing for growth in the
photic zone, generating a vertical gradient of chlorophyll.

\begin{figure}
\begin{centering}
\includegraphics[width=1\columnwidth]{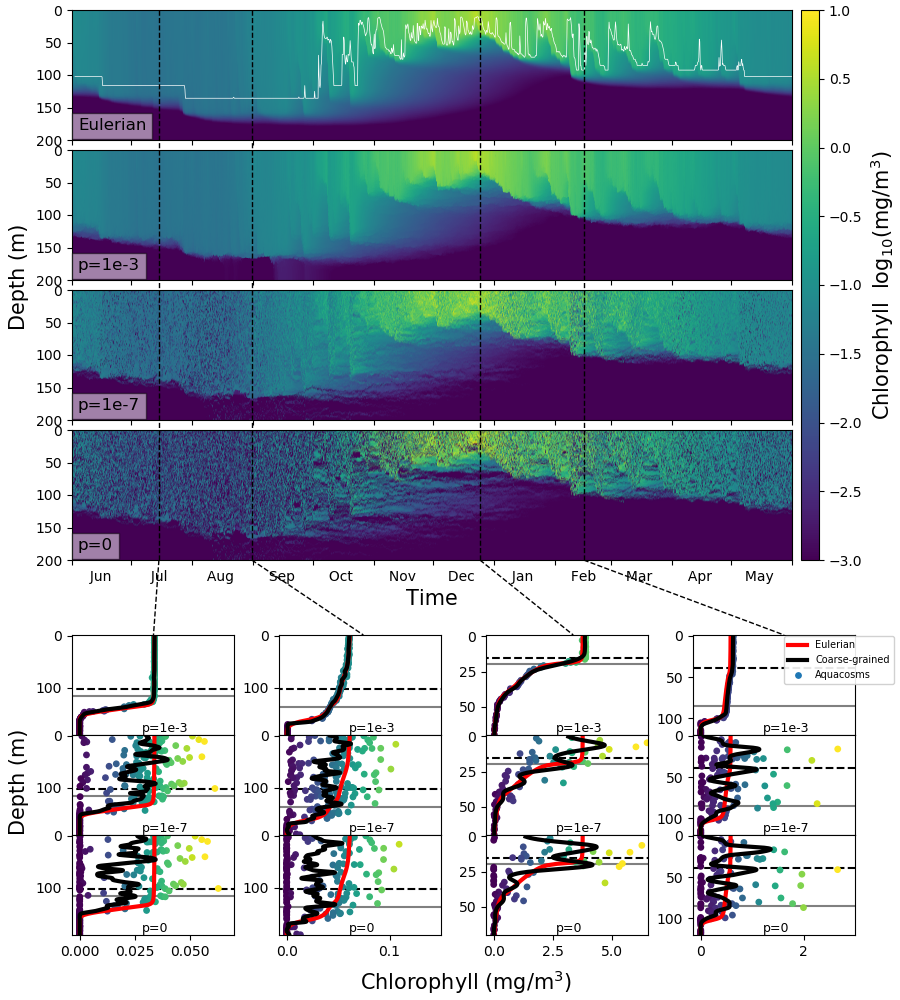}
\par\end{centering}
\caption{\label{fig:SAZ_chlorophyll}Top four panels: chlorophyll in the SAZ
simulation in logarithmic units. The top panel shows the eddy-diffusive
Eulerian model, and the thin white line marks the mixed layer depth;
the next three show the aquacosm simulations for $p=10^{-3},10^{-7},0$.
The lower panels show, as a function of depth, the chlorophyll content
in mg/m$^{3}$ of the aquacosms (dots) for different values of $p$,
their coarse-grained version (black line, see methods), the mixed
layer depth (horizontal gray line) and the turbocline depth (horizontal
black dashed line) at the date marked in the upper panels by the vertical
black dashed lines. For comparison, the chlorophyll concentration
as a function of depth computed with the Eulerian simulation (red
line) is repeated in all the panels corresponding to the same date.}
\end{figure}

In our models, slacking turbulence and vertical gradients of light
definitely produce vertical gradients of chlorophyll, even when stirring
is modeled as irreversible mixing (see e.g. the Eulerian simulation
profiles in in mid April and end of May in Fig. \ref{fig:PAPA_chlorophyll}),
and effects which we neglect, such as light-dependent grazing, may
greatly enhance these gradients \cite{moeller2019}. However, patchiness
at the microscale is the dominant source of variance in the mixed
layer. Patchiness is visually evident in the bottom panels of Fig.
\ref{fig:SAZ_chlorophyll}. At lower values of $p$, very large relative
differences in the chlorophyll content of nearby aquacosms are normal
even above the turbocline. This extreme variability is quantified
in Fig. \ref{fig:SAZ_coefficient_of_variation}A (see supplementary
Fig. \ref{fig:PAPA_coefficient_of_variation}A for the PAPA simulations),
showing the monthly average of the coefficient of variation of chlorophyll
(ratio of the standard deviation and the mean) computed above the
turbocline depth, thus excluding the effects of slacking turbulence.
In simulations with moderate and low irreversible mixing, the coefficient
of variation is never negligibly small, and even when the mixed layer
is deepest and the turbulence is strongest, it doesn't drop below
about 0.5. As expected, the variability is larger during the Austral
summer months, when turbulence is weaker than in other seasons, showing
an extended peak from spring to autumn, in full agreement with the
variability observed through autonomous observations in the SAZ \cite{little2018}.
In the Eulerian simulation, and in Lagrangian simulations with very
strong irreversible mixing, the peak is small and occurs in December,
when the mixed layer is the shallowest, while the coefficient of variation
remains negligibly different from zero in other months.

These results, for $p=10^{-6}$ or lower, bear a strong resemblance
to the statistics of the ARGO float observations in the Southern Ocean
(Fig. 6 in \cite{carranza2018}), but suggest that, rather than by
external forcing, chlorophyll variability is mostly caused by differences
in the Lagrangian histories of water parcels \cite{kida2017lagrangian,baudry2018}
but modulated by irreversible mixing. We remark that ARGO floats are
not high-resolution chlorophyll profilers \cite{carranza2018}, and
can't accurately represent vertical fluctuations on scales smaller
than a few meters. The black lines in the lower panels of Fig. \ref{fig:SAZ_chlorophyll}
have been computed from the aquacosm concentration using a smoothing
procedure yielding 5 m of resolution (see methods), thus comparable
with that of the floats. Chlorophyll fluctuations are damped, but
not completely wiped out. The resolution length scale of observations
affects variance estimations, as found when high-frequency sampling
instruments are used \cite{little2018}. In Figure \ref{fig:SAZ_coefficient_of_variation}B
(Fig. \ref{fig:PAPA_coefficient_of_variation}B for PAPA) we show
the monthly average of the coefficient of variation of the simulation
data relative to $p=10^{-7}$, after they underwent this coarse-graining
procedure at several resolutions. Extreme smoothing yields estimates
of fluctuations not far from those of the Eulerian simulations, which
progressively increase as the resolution increases. Thus, albeit the
ARGO data are surprising in the amount of variability that they show,
we suspect that this is still an underestimation of the reality.

\begin{figure}
\begin{centering}
\includegraphics[width=1\columnwidth]{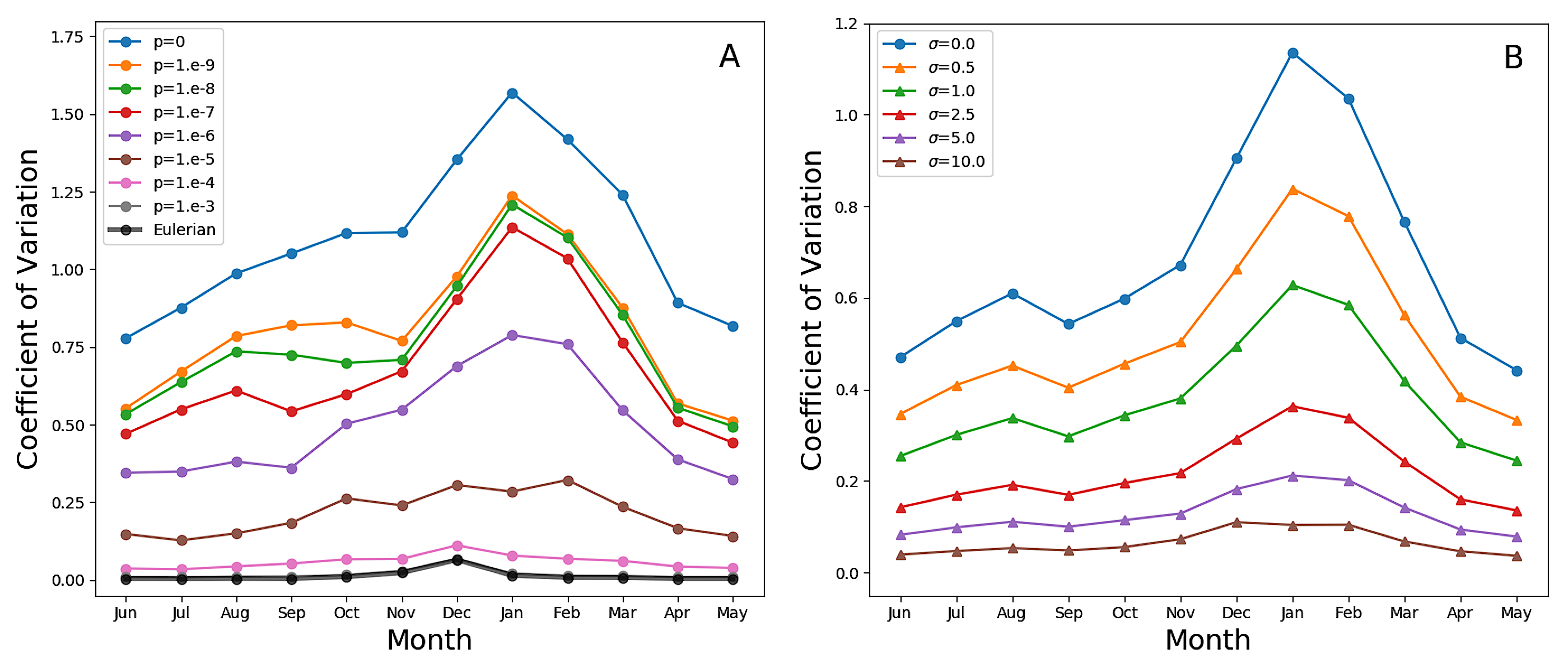}
\par\end{centering}
\caption{\label{fig:SAZ_coefficient_of_variation}\textbf{A}: monthly average
coefficient of variation (standard deviation over mean) of the chlorophyll
above the turbocline depth in the SAZ simulation. A value of 0 indicates
homogeneity, while a value of 1 implies departures from the vertical
mean that are as large as the mean itself. The thin lines in color
refer to the Lagrangian aquacosms with different values of the coupling
parameter $p$. The thick black line refers to the eddy-diffusive
Eulerian simulation. \textbf{B}: monthly average coefficient of variation
of the Lagrangian simulation with $p=10^{-7}$, and the same quantity
computed from profiles coarse-grained using a Gaussian kernel estimator
with standard deviation $\sigma=0.5,\,1,\,2.5,\,5,\,10$ m. The line
$\sigma=0$ refers to the uncoarsened results.}
\end{figure}

\section*{Discussion and Conclusion}

When considering mesoscale, and, more recently, submesoscale dynamics,
it has often been stressed that the joint effect of turbulent stirring
and nonlinear biochemical processes must produce an uneven, patchy
distribution of active tracers, and this, in turn, may affect the
bulk productivity and structure of oceanic ecosystems (see \cite{martin2003,levy2013,mahadevan2016,levy2018}
and references therein). Here we remark that the fundamental idea
expressed in those studies should also be scrutinized at smaller scales,
e.g. across the water column.

There exists a very large body of literature on the specific problem
which is the focus of this paper, namely the onset of open-ocean blooms.
Some of this literature tackles the problem of how different turbulence
properties affect biological growth. Yet the distinction between the
time scales of turbulent stirring and the time scales of irreversible
mixing is never considered. In spite of mounting evidence of ubiquitous
presence of patchiness in the vertical direction across the mixed
layer, theories of the onset of the bloom freely interchange turbulent
stirring and irreversible mixing as if they were one and the same
thing, (see the recent review \cite{fischer2014} and references therein).
On the other hand, the literature on mixed layer plankton patchiness
\cite{huisman2006,durham2011,cullen2015,moeller2019} focuses on unveiling
the underlying mechanisms but does not investigate how patchiness
contributes to signal at larger scales and how it should be included
in predictive models. In the present work we identified two simple
and distinct mechanisms that create patchiness vertically in the mixed
layer, and we showed how that affects longer time scales, such as
the phenology of the spring bloom. The first mechanism is essentially
physical: when deeper, phytoplankton-poor water is entrained by turbulence
into the phytoplankton-rich mixed layer, rapid mechanical stirring
produces a highly patchy water column. The second requires the existence
of a depth-dependent growth/decay process (e.g. due to the vertical
gradient of light) acting on time scales faster than the stirring
time scales. When this occurs, the uneven growth at different depths
creates a vertical gradient of the active tracers, which breaks down
into patchiness under the action of stirring.

Eulerian models that replace unresolved stirring with irreversible
mixing can't generate any patchiness from either of these mechanisms.
Because the time scale of removal of the fluctuations is the same
as that of stirring, (or, equivalently, because there is no way to
distinguish between eddy and molecular diffusion) fluctuations around
the local mean of phytoplankton and of any other tracer in the model
are wiped out with extreme efficiency (Figs. \ref{fig:stirring_vs_mixing},\ref{fig:Logistic},\ref{fig:Logistic_Sverdrup}).
Recent measurements have shown that these fluctuations are large in
the real ocean \cite{doubell2014}, and therefore the classical decomposition
methods, such as the one proposed by \cite{mandal2016}, are insufficient
to describe their nature. Lagrangian ensembles and individual-based
models, where transport is described by displacing the position of
the particles, produce patchiness through both mechanisms, but each
particle fully preserves its individual history (there is no equivalent
of molecular diffusion), so that its biochemical evolution remains
completely independent from that of all the others, to the point of
generating paradoxical results. 

We present a modeling approach which is Lagrangian in nature, but
allows for locally interacting particles; using the aquacosm concept,
growth, stirring and irreversible mixing can be treated separately,
as interacting but mutually independent parts of a complete biogeochemical
model. Because the reaction terms are representative of the biogeochemical
dynamics occurring in a very small, homogeneous water mass, they can
effectively include empirical reaction norms derived from laboratory
experiments and can retain the relationship with the environmental
drivers as they were originally measured \cite{boyd2018}. In other
words, our approach simulates biogeochemistry at the sub-microscale,
and therefore doesn't need to deal with ``effective'' biological
parameters or other bulk formulae as in the case of Eulerian models,
contributing to unambiguously identifying the mechanistic processes
which give rise to the complex phenomena occurring in plankton ecosystems.
Irreversible mixing, modeled as mass fluxes between nearby aquacosms,
connects the microscopic scales of the biogeochemical processes with
the macroscopic ones of the physical stirring, cutting through the
unresolved scales. 

Aquacosm simulations are just as under resolved as Eulerian simulations
having a number of mesh nodes comparable to the number of Lagrangian
particles, but they treat unresolved turbulent stirring and irreversible
mixing as separate and distinct processes. The time scale associated
to the latter is set through the choice of the parameter $p$. Formally,
as specified by eq. (\ref{eq:exchange_fractions}) in ``methods'',
$p$ determines the amount of the material oozed out of the $i-$th
aquacosm that ends up being caught into the $j-$th within a time
step. In a three-dimensional setting, this number must be interpreted
as a volume (e.g., $p=10^{-6}$ in a simulation using meters as unit
of length would correspond to aquacosms of one cubic centimeter of
volume). A better interpretation of the significance of $p$ and of
the other parameters determining the strength of the irreversible
mixing (interparticle distance, time step, etc.) is developed in the
supplementary material S3. There we show that the coupling between
aquacosms (eq. (\ref{eq:coupling}) in ``methods'') can be associated
to a diffusion coefficient, which in aquacosms simulations plays the
role of molecular diffusion. We find that with the PAPA simulation
parameters and $p$ in the range between $10^{-7}$ and $10^{-8}$,
this coefficient assumes a value of the same order of magnitude of
the diffusivity of seawater ions, thus suggesting that this is a realistic
range for those simulations.

We find that, depending on the degree of irreversible mixing, the
onset of the bloom is shifted by a number of days comparable with
the shifts that may occur at the end of the century \cite{henson2018}
according to climatological models. For strengths of irreversible
mixing that we consider realistic, we find a shift which would largely
mitigate the problem of anticipated blooms in the Southern Ocean \cite{hague2018}
plaguing current Eulerian models. This and other stubborn biases have
often been attributed to inadequacies of the biological formulation,
but are likely to stem from mismodeling the interaction, across vastly
different scales, of growth and stirring, mediated by irreversible
mixing \cite{mckiver2015impact}.

Aquacosm simulations offer an ideal tool for exploring which biological
features are able to build up large-scale impacts, and which are negligible
in terms of bulk properties. The aquacosm approach is not limited
to the extremely simplified treatment of the growth/decay processes
that we have used here to illustrate the potentialities of the method,
and can be expanded to include all the biogeochemical processes that
may be deemed relevant for the specific problem at hand.

\section*{Methods}

\subsection*{The aquacosm approach}

A generic Lagrangian-ensemble water-column model is embodied by the
following equations
\begin{equation}
dz_{i}=\frac{\partial}{\partial z}\kappa(z_{i},t)\,dt+\sqrt{2\kappa(z_{i},t)}\,dW_{i}\label{eq:Brownian_motion}
\end{equation}
\begin{equation}
\begin{cases}
\dot{c}_{i}^{(1)} & =f_{1}(c_{i}^{(1)},\ldots,c_{i}^{(m)},z_{i},t)\\
 & \;\vdots\\
\dot{c}_{i}^{(m)} & =f_{m}(c_{i}^{(1)},\ldots,c_{i}^{(m)},z_{i},t)
\end{cases}\label{eq:reaction_kinetics}
\end{equation}
where the index $i=1,\ldots,N$ identifies the particle having depth
$z_{i}$, that performs a Brownian motion characterized by an eddy
diffusivity $\kappa$, which may depend on depth and time $t$; $W_{i}$
is a realization of the standard Wiener process. See \cite{grawe2011implementation,van2018lagrangian}
for a derivation of equation (\ref{eq:Brownian_motion}). The quantities
$c_{i}^{(1)},\ldots,c_{i}^{(m)}$ are the concentrations of the $m$
scalar quantities describing the planktonic ecosystem (e.g. in an
NPZD model it would be $m=4$, with concentrations of nutrient, phytoplankton,
zooplankton and detritus, respectively) and the overlying dot denotes
the time derivative. The functions $f_{1},\ldots,f_{m}$ describe
the reaction kinetics, where the dependence on depth and time accounts
for the effect of light and its daily and seasonal variations, and
for any other external forcing.

In the aquacosm approach we interpret the Lagrangian particles as
tiny control volumes. They should be thought of as minuscule aquatic
mesocosms carried by the ocean dynamics, and which are homogeneous
in their scalar content. This interpretation is shared with the Lagrangian-ensemble
models, but, to avoid the issues discussed in the main text, we propose,
in addition, to allow mass exchanges between nearby aquacosms. We
define the mass fraction $q_{ij}$ that the $i-$th aquacosm gives
to the $j-$th aquacosm as 
\begin{equation}
q_{ij}=\begin{cases}
p\left(4\pi K_{ij}\Delta t\right)^{-\nicefrac{1}{2}}\exp\left(-\frac{\left|z_{i}-z_{j}\right|^{2}}{4K_{ij}\Delta t}\right), & \left|z_{i}-z_{j}\right|<R\\
0, & \left|z_{i}-z_{j}\right|\ge R
\end{cases}\label{eq:exchange_fractions}
\end{equation}
then, at intervals of time $\Delta t$, we update the concentrations
carried by each particle as
\begin{equation}
c_{i}^{(l)}\longleftarrow c_{i}^{(l)}-\sum_{j=1}^{N}q_{ij}c_{i}^{(l)}+\sum_{j=1}^{N}q_{ji}c_{j}^{(l)},\quad l=1,\ldots,m\label{eq:coupling}
\end{equation}
for all the scalars $l=1,\ldots,m$. Here the first sum represents
the mass fraction that leaves the $i-$th aquacosm and is redistributed
to all the other aquacosms, and the second sum, conversely, represents
an equal mass fraction received by the $i-$th aquacosm from all the
others (note that $q_{ij}=q_{ji}$). The received mass fraction is
composed of many distinct parts, each carrying the concentration of
the scalars contained in the acquacosm of provenance. These parts
immediately and irreversibly homogenize with the remaining content
of the $i-$th aquacosm in order to determine its new concentration
values. Here $p$ is a free parameter, which in this one-dimensional
formulation has the dimensions of a length, but in three dimensions
would be a volume, that can be used to tune the coupling strength
between the aquacosms (choosing $p=0$ is equivalent to using the
a Lagrangian ensemble model with uncoupled particles). The variance
of the Gaussian kernel coupling the $i-$th and $j-$th aquacosms
is chosen on the basis of the eddy diffusion coefficient as 
\[
K_{ij}=\min\left\{ \kappa(z_{i},t),\;\kappa(z_{j},t)\right\} .
\]
In order to allow for an efficient numerical implementation, the coupling
between aquacosms further apart than some threshold distance $R$
must be zero. This algorithm conserves mass and avoids the creation
of spurious maxima and minima \cite{paparella2018}, provided that
the parameters are chosen so that 
\[
\forall i,\quad\sum_{j=1}^{N}q_{ij}\le1.
\]

Roughly speaking, one may think the aquacosms as oozing out part of
their content into Gaussian clouds spreading at a rate specified by
the eddy diffusion coefficient. Then, at regular intervals of time
$\Delta t$, all the material within a control volume (both what was
left inside the volume, and what came from the overlapping clouds)
is instantaneously and irreversibly homogenized, thus determining
the concentrations which will evolve according to (\ref{eq:reaction_kinetics})
for the next interval of time $\Delta t$. A conceptually similar
technique describing advection-diffusion processes as the interleaving
of short time intervals of pure transport alternated with instantaneous
irreversible mixing events has already been successfully used to model
mixed layer dynamics \cite{ferrari1997development}. These sort of
modeling procedures have their justification rooted in the fractional
step method for the numerical solution of differential equations.

All Lagrangian results presented in this paper use 200 aquacosms.
Equations (\ref{eq:Brownian_motion}), (\ref{eq:reaction_kinetics})
were integrated \cite{grawe2011implementation,van2018lagrangian}
using Milstein's and the midpoint methods, respectively, with a time
step $\Delta t=10^{-5}$ non-dimensional time units for the idealized
cases of Figures \ref{fig:stirring_vs_mixing}-\ref{fig:Logistic_Sverdrup}
and $\Delta t=5$ s for the PAPA and SAZ cases. The eddy diffusivity
profiles for the latter cases were generated by a physical ocean model
described below and were interpolated at the position of the aquacosms
with monotone B-splines \cite{ross2004recipe}. For simplicity, reflecting
boundary conditions were imposed at 200 m of depth. The interaction
radius was $R=0.05$ non–dimensional units for the idealized cases
and $R=10$ m for the PAPA and SAZ simulations.

Coarse-grained profiles were obtained by smoothing the concentrations
with a Gaussian kernel estimator \cite{grawe2011implementation} having
a standard deviation of $1/20$ of the domain for the idealized cases
and 2.5 m for PAPA and SAZ, or as otherwise specified in the figure
caption.

\subsection*{Eulerian models}

All the Eulerian models in this paper are solved with an explicit,
second-order finite differences scheme, with $C$ and $\kappa$ evaluated
on staggered uniform grids. The simulated water column is one non-dimensional
length unit with mesh size of 1/200 units for the idealized cases
and 200 m deep for PAPA and SAZ, with a mesh size of 1 m (see below).
In all cases no-flux boundary conditions are imposed at the top and
bottom of the water column. The eddy diffusivity $\kappa$ is interpolated
onto the uniform grid with the same B-spline interpolator used for
the Lagrangian simulations.

\subsection*{The PAPA and SAZ models}

The chosen station locations are representative of two typical stratification
regimes in the open ocean. Since we focused on the relationship between
turbulence and light, the key distinguishing feature is the time evolution
of the vertical water column structure. Weather station PAPA is located
in the North-East Pacific (50\textdegree N, 145\textdegree W), and
it is characterized by mixing confined to less than 100 m with maximum
cooling in March-April and the development of summer stratification
between June and October. The PAPA station has been used in the literature
to develop and analyze turbulence closure models \cite{burchard01,reffray15}.
We used a 1-D version of the NEMO physical ocean model with the parameterizations
described in \cite{reffray15}. The model was run with 75 vertical
levels and forced by ECMWF ERA-interim reanalyses \cite{dee2011},
to obtain the hourly values of eddy diffusivity used in the Eulerian
and Lagrangian biogeochemical models. A similar model was developed
for the Sub-Antarctic zone of the Southern Ocean (SAZ), using the
same vertical grid and same type of atmospheric forcing as in PAPA.
This model site is ideally located in the Atlantic sector at 45\textdegree S
8\textdegree E, in similar light conditions as for PAPA. This region
features deep mixing beyond 100 m between May and August and weak
stratification during the Austral summer months.

To illustrate the versatility of the aquacosm approach to incorporate
any kind of biogeochemical model with varying complexity beyond the
non-dimensional study cases, we used a simplified version of the Biogeochemical
Flux Model \cite{vichi2007}. The chosen formulation tracks phytoplankton
carbon concentration $C$, measured in mg m$^{-3}$ for a generic
functional type of mid-sized diatoms, in which growth is only limited
by light availability and an implicit temperature dependence is included
in the parameter choice to account for the different oceanic regions.

The photosynthetic available radiation $E_{_{PAR}}$is propagated
according to the Lambert-Beer formulation 
\begin{equation}
E_{_{PAR}}(z)=\varepsilon_{_{PAR}}\,\,Q_{S}\,e^{\lambda_{w}z+\int_{z}^{0}\lambda_{bio}(z')dz'}\label{eq:irradiance_bio}
\end{equation}
where $Q_{s}$ is the net broadband solar radiation at the surface
from ERA-interim (W m$^{-2}$), $\varepsilon_{_{PAR}}=0.4/0.217$
is the coefficient determining the fraction of photosynthetically-available
radiation (converted to $\mu$E m$^{-2}$s$^{-1}$ using the constant
0.217). Light propagation takes into account the extinction due to
pure water $\lambda_{w}$ (0.0435 m$^{-1}$) and to phytoplankton
concentration $\lambda_{bio}$. The broadband biological light extinction
is approximated to a linear function of the phytoplankton chlorophyll
concentration $L$
\begin{equation}
\lambda_{bio}=cL\label{eq:bio_extinction_coefficient}
\end{equation}
regulated by the specific absorption coefficient (c = 0.03 m$^{2}$
mg chl$^{-1}$). To be more comparable with the non-dimensional idealized
experiments, this very simple model neglects photoacclimation phenomena,
therefore we assume 
\begin{equation}
L=\theta_{chl}C\label{eq:chlorophill_conversion}
\end{equation}
where the chlorophyll to carbon ratio $\theta_{chl}$ was taken to
be $0.017$ mg \emph{chl} mg \emph{C}$^{-1}$ for PAPA and $0.013$
for SAZ \cite{behrenfeld2005,thomalla2017}. The same results (not
shown) were confirmed using the BFM acclimation model with variable
chlorophyll, based on the Geider et al. formulation \cite{geider1997,vichi2007}.
The carbon concentration rate of change is controlled by gross primary
production, respiration and a crowding mortality term that parameterizes
zooplankton grazing:
\begin{equation}
\dot{C}=rf^{E}C-bC-\frac{aC^{2}}{C_{h}+C}\label{eq:Simple_BFM}
\end{equation}
where $r$ is the maximum specific photosynthetic rate, $b$ is the
basal specific respiration rate, $a$ is the specific crowding mortality
rate and $C_{h}$ is the crowding half-saturation. Owing to the difference
in the seasonal cycle of nutrients and water temperature, we use $r=2$,
$b=0.16$, $a=1$ days$^{-1}$ for PAPA and $r=0.5$, $b=0.04$, $a=0.25$
days$^{-1}$ for SAZ. The lower potential growth rate in SAZ is derived
by applying a Q10 relationship \cite{vichi2007} and considering the
mean temperature during the bloom period. The other parameters are
tuned to yield realistic values of chlorophyll at the study sites.
In all cases we set $C_{h}=12.5$ mg m$^{-3}$. The light regulating
factor is defined as
\begin{equation}
f^{E}=1-\exp\left(-\frac{\alpha E_{PAR}}{r}\theta_{chl}\right)\label{eq:light-regfac}
\end{equation}
where $\alpha=1.38\,10^{-5}$ $\mu$E$^{-1}$m$^{2}$ \cite{vichi2007}.

The eddy-diffusive Eulerian version of this model is
\begin{equation}
\frac{\partial C}{\partial t}=\frac{\partial}{\partial z}\left(\kappa(z,t)\frac{\partial C}{\partial z}\right)+\dot{C}.\label{eq:Eulerian_PAPA_SAZ}
\end{equation}
The Lagrangian aquacosm models use $m=1$ and (\ref{eq:reaction_kinetics})
reduces to (\ref{eq:Simple_BFM}).

Starting from initial conditions having a small constant concentration,
the runs extend for four years (each year repeats the same eddy diffusivity
and radiation data). Except for the first year, in all models the
results have negligible differences between the years.

\bibliographystyle{naturemag}
\bibliography{aquacosm}

\pagebreak{}

\section*{Supplementary materials}

\setcounter{equation}{0} 
\renewcommand{\theequation}{S\arabic{equation}}

\subsection*{S1: On the equivalence of the Eulerian and Lagrangian approaches
in Sverdrup's theory.}

When the concentration $C\ll1$ the equation (\ref{eq:Logistic_non-dimensional})
may be linearized to
\begin{equation}
\frac{\partial C}{\partial t}=\varepsilon f(z)C+\frac{\partial^{2}C}{\partial z^{2}}\label{eq:Sverdrup_non-dimensional}
\end{equation}
Following Sverdrup, we consider the case of very vigorous stirring,
that is $\varepsilon\ll1$. Then, because the diffusion term is the
dominant one, one contends that the plankton field $C$ will be nearly
independent of $z$, except, possibly, for an initial transient lasting
no more than the mixing time scale. Thus, on integrating the above
equation over the vertical domain, and exploiting the no-flux boundary
conditions in the diffusion term, the time evolution of the water-column
averaged phytoplankton concentration $\left\langle C\right\rangle $
is found to evolve approximately according to 
\begin{equation}
\frac{d}{dt}\left\langle C\right\rangle =\varepsilon I\left\langle C\right\rangle \label{eq:Sverdrup_average_evolution}
\end{equation}
whose solution is
\begin{equation}
\left\langle C\right\rangle (t)=\left\langle C\right\rangle (0)\,e^{\varepsilon It}.\label{eq:Sverdrup_solution}
\end{equation}
The sign of the constant 
\begin{equation}
I=\int_{0}^{1}f(z)\,dz\label{eq:I}
\end{equation}
determines whether the plankton population, overall, will grow or
decay. Using Sverdrup's form (\ref{eq:Sverdrup_f}) for $f$, one
readily recognizes that growth is possible only if the mixed layer
depth $\ell$ is not too deep with respect to the length scale $\lambda$.

In Lagrangian-ensemble models, a fluid parcel having depth $z_{i}$
carries a homogeneous concentration $C_{i}$ of phytoplankton. The
trajectories of the fluid particles are generally modeled as sample
paths of a Brownian motion. Regardless of the details of how their
trajectory is modeled, a really important underlying hypothesis (most
often not explicitly stated) is that the following ergodic identity
holds
\begin{equation}
\lim_{t\to\infty}\frac{1}{t}\int_{0}^{t}f(z_{i}(\tau))\,d\tau=\int_{0}^{1}f(z)\,dz\label{eq:ergodic}
\end{equation}
where the identity must be valid for any possible choice of $f$ and
for all fluid parcels $z_{i}$ (except, at most, for a set of measure
zero). Namely, the time average of the value of $f$ experienced by
the typical fluid parcel along its trajectory must converge to the
spatial average of $f$, which is the constant $I$ appearing in (\ref{eq:Sverdrup_average_evolution}).
The concentration of plankton carried by each particle is assumed
to evolve according to the same reaction kinetics as in the Eulerian
case. This leads to the following equation for each fluid parcel
\[
\frac{d}{dt}C_{i}=\varepsilon f(z_{i}(t))\,C_{i}.
\]
 Separating the variables, and integrating in time, we have
\begin{equation}
\int_{C_{i}(0)}^{Ci(t)}\frac{dC_{i}}{C_{i}}=\varepsilon\int_{0}^{t}f(z_{i}(\tau))\,d\tau\label{eq:Lagrangian separation of variables}
\end{equation}
If the stirring time scale, as quantified by the eddy diffusion coefficient,
is much shorter than the physiological phytoplankton growth time scale,
then one may contend that the ergodic identity (\ref{eq:ergodic})
will be approximately true not just in the limit $t\to\infty$, but
also after any finite time $t$ longer than the mixing time scale.
Therefore, one substitutes the integral in the right-hand side of
(\ref{eq:Lagrangian separation of variables}) with $I$ and finds
the following approximate solution
\begin{equation}
C_{i}(t)=C_{i}(0)e^{\varepsilon It}\label{eq:Lagrangian}
\end{equation}
If one tracks $N$ Lagrangian particles, uniformly seeded along the
water column, then the arithmetic average of their plankton concentration
\[
\overline{C}^{N}(t)\equiv\frac{1}{N}\sum_{i=1}^{N}C_{i}(t)=\overline{C}^{N}(0)\,e^{\varepsilon It}
\]
will converge to Sverdrup's Eulerian solution (\ref{eq:Sverdrup_solution})
in the limit $N\to\infty$.

Even though in Sverdrup's model the Lagrangian and Eulerian approaches
yield identical results, this does not generalize to nonlinear cases.
For instance, if we use the logistic growth model
\begin{equation}
\frac{\partial C}{\partial t}=\varepsilon f(z)C\left(1-C\right)+\frac{\partial^{2}C}{\partial z^{2}}\label{eq:Eulerian_logistic_non-dimensional}
\end{equation}
with the same hypothesis and approximations as in Sverdrup's model,
we find the Eulerian solution
\begin{equation}
\left\langle C\right\rangle (t)=\frac{\left\langle C\right\rangle (0)e^{\epsilon It}}{\left\langle C\right\rangle (0)\left(e^{\epsilon It}-1\right)+1}\label{eq:Logistic_Eulerian_solution}
\end{equation}
where $\left\langle C\right\rangle (0)$ is the average plankton concentration
of the initial condition. The Lagrangian model yields, along each
particle
\begin{equation}
C_{i}(t)=\frac{C_{i}(0)\,e^{\epsilon It}}{C_{i}(0)\left(e^{\epsilon It}-1\right)+1}.\label{eq:Logistic_Lagrangian_solution}
\end{equation}
The arithmetic average of (\ref{eq:Logistic_Lagrangian_solution})
over $N$ Lagrangian particles does not converge to the Eulerian solution
(\ref{eq:Logistic_Eulerian_solution}) in the limit $N\to\infty$.
For example, take an initial condition containing only, and in equal
proportions, fluid parcels with either $C_{i}(0)=1$ or $C_{i}(0)=0$.
Then, the arithmetic average of (\ref{eq:Logistic_Lagrangian_solution})
is:
\begin{equation}
\overline{C}^{N}(t)=\frac{1}{2}.\label{eq:Logistic_Lagrangian_average}
\end{equation}
With the Eulerian approach, starting from $\left\langle C\right\rangle (0)=\nicefrac{1}{2}$,
(\ref{eq:Logistic_Eulerian_solution}) describes growing population
that reaches $\left\langle C\right\rangle =1$ asymptotically in time.
The Lagrangian and the Eulerian approaches give irreconcilably different
results.

\subsection*{S2: Growth rates in weak turbulence}

Sverdrup's model yields an accurate approximation of the solution
of equation (\ref{eq:Sverdrup_non-dimensional}) when $\varepsilon$
is no larger than one, that is, when the reaction time scales are
no faster than the stirring time scales, as quantified by the eddy
diffusion coefficient. In the opposite case, it is inappropriate to
assume that turbulence is able to maintain a nearly constant concentration
of plankton in the water column: plankton will be more abundant close
to the surface than at depth, and this results in a faster growth
rate than predicted by Sverdrup's theory. In order to estimate the
growth rate, we seek solutions of (\ref{eq:Sverdrup_non-dimensional})
of the form 
\[
C(z,t)=A_{n}(t)\phi_{n}(z).
\]
Separating the variables, one obtains: 
\[
A_{n}(t)=A_{n}(0)\exp(\sigma_{n}t)
\]
where the growth rate $\sigma_{n}$ is determined by seeking a non-zero
solution $\phi_{n}$ of the eigenvalue problem 
\begin{equation}
\phi_{n}^{\prime\prime}+\varepsilon f\phi_{n}=\sigma_{n}\phi_{n}\label{eq:Sturm-Liouville problem}
\end{equation}
where $\phi_{n}$ is subject to the same boundary conditions as $C$
(no-flux in our case). Sturm-Liouville theory (see, e.g. \cite{strauss2007})
insures the existence of a countably infinite set of pairs $\left(\sigma_{n},\phi_{n}\right)$,
with $\sigma_{n}$ ordered and decreasing with $n$. The Rayleigh
quotient associated to the problem (\ref{eq:Sturm-Liouville problem})
is 
\begin{equation}
q(\psi)=\frac{\varepsilon\int_{0}^{1}f\psi^{2}dz-\int_{0}^{1}\left(\psi^{\prime}\right)^{2}dz}{\int_{0}^{1}\psi^{2}dz}\label{eq:Rayleigh quotient}
\end{equation}
On multiplying (\ref{eq:Sturm-Liouville problem}) by $\phi_{n}$
and integrating by parts, it is easy to verify that setting $\psi=\phi_{n}$
gives $q(\psi)=\sigma_{n}$. Among all non-zero, differentiable functions
$\psi$, the one which maximizes the quotient is $\phi_{0}$. A simple,
explicit form for $\phi_{0}$ can not, in general, be obtained. However,
using a physically motivated choice for $\psi$ in the Rayleigh quotient,
we can still seek an approximation from below of the value of $\sigma_{0}$,
which is the growth rate that we expect to observe in a solution of
equation (\ref{eq:Sverdrup_non-dimensional}) starting from initial
conditions very close to zero.

Sverdrup's theory uses the approximation $\phi_{0}(z)\approx\psi(z)=1$,
which yields the following estimate for the growth rate
\[
\sigma_{0}\approx q(\psi)=\varepsilon\int_{0}^{1}f(z)\,dz
\]
as in equation (\ref{eq:Sverdrup_non-dimensional}). This is an appropriate
maximization strategy for small $\varepsilon$: when the first integral
at the numerator of (\ref{eq:Rayleigh quotient}) is multiplied by
a tiny number, then the second integral must be kept small by maintaining
$\psi^{\prime}$ very close to zero in order to obtain the largest
possible value of $q$. In the case of weak turbulence, that is, for
large $\varepsilon$, the plankton residing in the sunlit region close
to the surface has time to grow before being fluxed down in the dark
depths. Therefore, the presence of a vertical plankton concentration
gradient is expected, and a functional form such as 
\begin{equation}
\psi_{\alpha}=1+\alpha\cos(\pi z)\label{eq:psi_alpha}
\end{equation}
will give a better approximation than Sverdrup's, if the value of
the constant $\alpha$ is chosen appropriately. Using the \emph{ansatz}
(\ref{eq:psi_alpha}) in (\ref{eq:Rayleigh quotient}) results in
the following quotient
\begin{equation}
q(\psi_{\alpha})=\frac{\varepsilon\left(2I+4\alpha I_{1}+2\alpha^{2}I_{2}\right)-\pi^{2}\alpha^{2}}{\alpha^{2}+2}\label{eq:Rayleigh quotient psialpha}
\end{equation}
where the Sverdrup integral $I$ is given in (\ref{eq:I}), and we
defined
\[
I_{1}=\int_{0}^{1}f(z)\cos(\pi z)\,dz;\quad I_{2}=\int_{0}^{1}f(z)\cos^{2}(\pi z)\,dz.
\]
As is verifiable by analytical means, the value of $q$ in (\ref{eq:Rayleigh quotient psialpha})
has a single maximum attained for positive $\alpha$. For small $\varepsilon$
such a maximum is nearly indistinguishable from the value attained
for $\alpha=0$, but, as $\varepsilon$ increases, the difference
becomes sizable. In particular, for $\epsilon=10$ we have $q(\psi_{0})=0.498\cdots$,
and $q(\psi_{\alpha_{max}})=0.820\cdots$. The latter value is the
growth rate used to plot the thin solid line in Figure \ref{fig:Logistic_Sverdrup},
showing, as expected, just a slight underestimation of the growth
rates observed both in the Eulerian and in the Lagrangian simulations,
in the linear regime.

\subsection*{S3: Strength of the irreversible mixing}

In order to evaluate the effects of the coupling formula (\ref{eq:coupling}),
so as to give guidance in choosing a realistic value for the parameter
$p$, it is convenient to consider the hypothetical case in which
the eddy diffusivity is a constant $K$, and the aquacosms are arranged
along the water column at equally spaced depths $z_{i}=ih$. If one
chooses an interaction length $R$ such that the $i-$th aquacosm
interacts only with the $(i-1)-$th and the $(i+1)-$th then, (\ref{eq:coupling})
in the main text becomes
\[
c^{(l)}(z_{i},t+\Delta t)=c^{(l)}(z_{i},t)-2qc^{(l)}(z_{i},t)+qc^{(l)}(z_{i+1},t)+qc^{(l)}(z_{i-1},t)
\]
where
\[
q=\frac{p}{\left(4\pi K\Delta t\right)}\exp\left(-\frac{h^{2}}{4K\Delta t}\right).
\]
This can be re-arranged as
\[
\frac{c^{(l)}(z_{i},t+\Delta t)-c^{(l)}(z_{i},t)}{\Delta t}=\frac{h^{2}}{\Delta t}q\left(\frac{c^{(l)}(z_{i+1},t)+c^{(l)}(z_{i-1},t)-2c^{(l)}(z_{i},t)}{h^{2}}\right)
\]
which is the forward in time, centered in space finite difference
method for the diffusion equation associated to the diffusion coefficient
\begin{equation}
\mathcal{D}=\frac{h^{2}}{\Delta t}q.\label{eq:molecular_diffusion}
\end{equation}
Because equation (\ref{eq:coupling}) represents the irreversible
mixing processes, it is legitimate to interpret $\mathcal{D}$ as
a molecular diffusion coefficient. For actual simulations where the
particles are scattered at uniformly random depths, the expression
(\ref{eq:molecular_diffusion}) can still be used as a guidance, provided
that $h$ is interpreted as the mean distance between first neighbors.
When the value of $R$ is such that more than three aquacosms interact
simultaneously, it is natural to expect a somewhat larger diffusivity,
because each aquacosm exchanges more mass with other aquacosms, but
if $R>\sqrt{2K\Delta t}$, this effect will be less than proportional
to the number of particles, because aquacosms substantially further
away than $\sqrt{2K\Delta t}$ contribute very little \cite{paparella2018}.

For the PAPA simulations $K\approx0.025$ m$^{2}$s$^{-1}$, $h=1$
m, $\Delta t=5$ s. We also have $R=10$ m, so that, on average, 20
aquacoms are simultaneously interacting, and $\sqrt{2K\Delta t}=0.5$
m. On choosing $p=10^{-7}$ we obtain $\mathcal{D}\approx2\cdot10^{-9}$
m$^{2}$s$^{-1}$. Considering that the molecular diffusivity of ions
in seawater is of the order of $10^{-9}$ m$^{2}$s$^{-1}$, it seems
reasonable to assume that realistic values for $p$ should range between
$10^{-8}$ and $10^{-7}$.

\renewcommand{\refname}{Supplementary References}

\section*{Supplementary figures}

\setcounter{figure}{0} 
\renewcommand{\thefigure}{S\arabic{figure}}

\begin{figure}
\begin{centering}
\includegraphics[width=1\columnwidth]{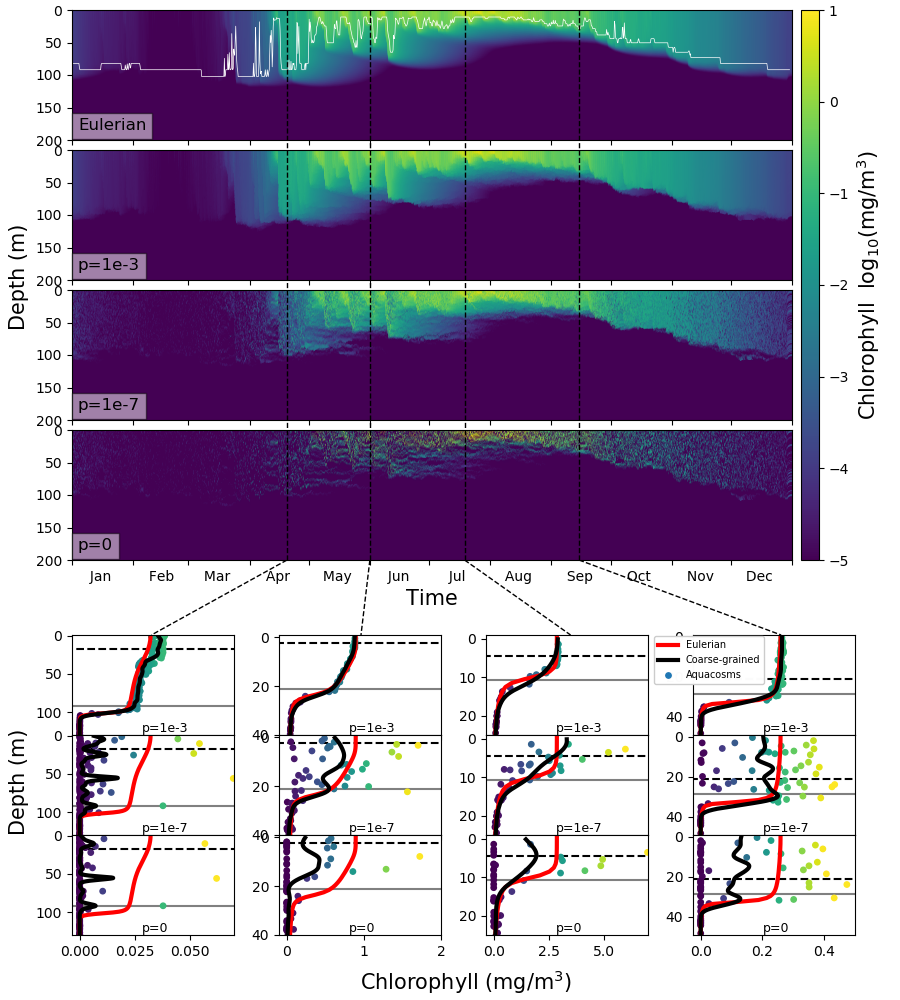}
\par\end{centering}
\caption{\label{fig:PAPA_chlorophyll}As Figure \ref{fig:SAZ_chlorophyll},
but for the PAPA simulations.}
\end{figure}

\begin{figure}
\begin{centering}
\includegraphics[width=1\columnwidth]{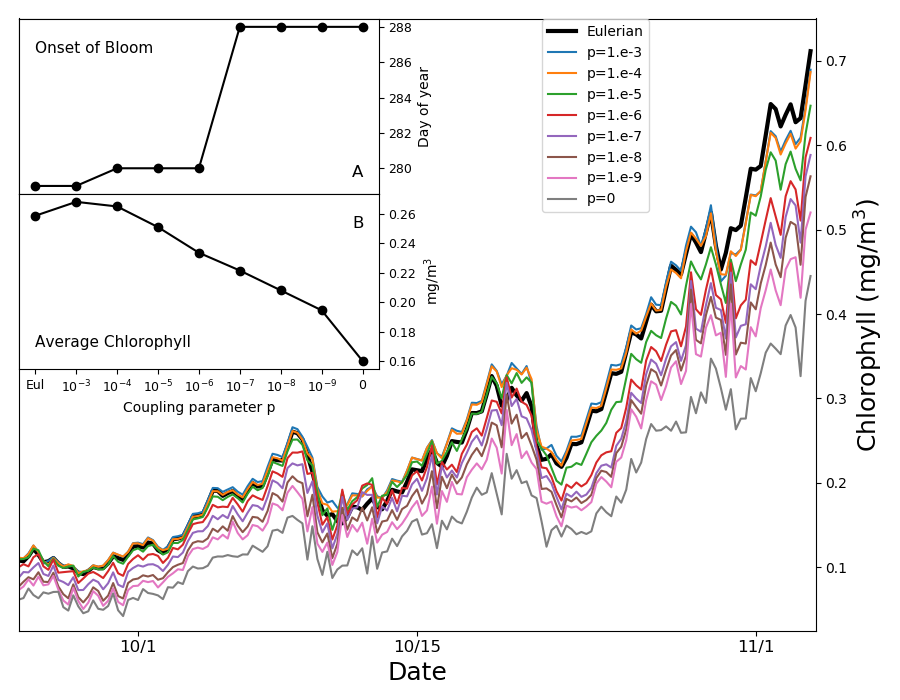}
\par\end{centering}
\caption{\label{fig:SAZ_phenology}As figure \ref{fig:PAPA_phenology}, but
for the SAZ simulations.}
\end{figure}

\begin{figure}
\begin{centering}
\includegraphics[width=1\columnwidth]{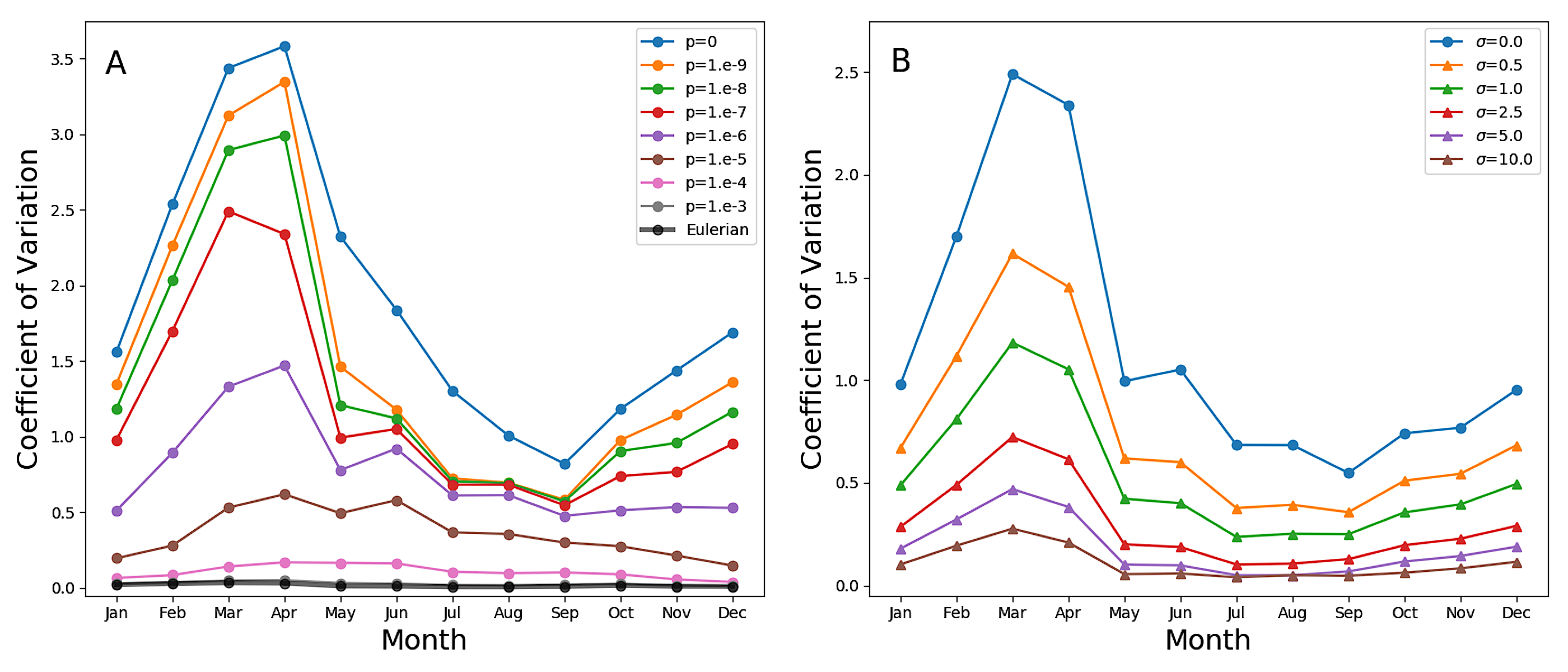}
\par\end{centering}
\caption{\label{fig:PAPA_coefficient_of_variation}As Figure \ref{fig:SAZ_coefficient_of_variation},
but for the PAPA simulations.}
\end{figure}

\end{document}